\documentstyle[12pt]{article}
\def\gappeq{\mathrel{\rlap {\raise.5ex\hbox{$>$}}
{\lower.5ex\hbox{$\sim$}}}}

\def\lappeq{\mathrel{\rlap{\raise.5ex\hbox{$<$}}
{\lower.5ex\hbox{$\sim$}}}}

\begin{document}
\topmargin -1.0cm
\oddsidemargin -0.8cm
\evensidemargin -0.8cm
\vspace*{2cm}
\begin{flushleft}
{\bf BEYOND THE STANDARD MODEL}\\
Gian F. Giudice\\
CERN, Geneva, Switzerland
\vspace{2cm}
\end{flushleft}

{\rightskip=5pc
\leftskip=5pc
\noindent
{\bf Abstract}\\
In these lectures I give a short review of the main
theoretical ideas underlying the extensions of the Standard Model
of elementary particle interactions.
\vglue 2.0cm}

\section{INTRODUCTION}
Studying physics beyond the Standard Model means looking for the
conditions of the
Universe in the first billionth of a second, when its temperature was
above
$10^{14}$ K.  This clearly requires a gigantic intellectual leap in
the investigation.  It is even more striking that modern
accelerators can
reproduce particle collisions similar to those that continually occurred
in the
thermal bath in the very first instants of our Universe.  We are
now
entering the age in which, with the joint effort of experiments and
theory, we
are likely to unravel the mystery of the fundamental principles of
particle
interactions lying beyond the Standard Model.

The Standard Model \cite{gws} 
describes the interactions of three generations of
quarks
and leptons defined by a non-Abelian gauge theory based on the group
$SU_3 \times SU_2 \times U_1$.  The precision measurements at LEP have given
an
extraordinary confirmation of the validity of the Standard Model up to
the
electroweak energy scale (for reviews, see ref.~\cite{ew}), and we have
no
firm experimental indications for failures of this theory at higher
energies.
Our belief that the Standard Model is a low-energy approximation of a
new and
fundamental theory is based only on theoretical, but well-motivated,
arguments.

First of all, the electroweak symmetry breaking sector is not on firm
experimental ground.  The Higgs mechanism, which is invoked by the
Standard
Model to generate the $Z^0$ and $W^\pm$ masses, predicts the existence
of a new
scalar particle, still to be discovered.  From the theoretical point of
view,
the Higgs mechanism suffers from the so-called ``hierarchy" or
``naturalness"
problem which, as discussed in sect.~5,
leads us to believe that new
physics
must take place at the TeV energy scale.

Furthermore, the complexity of the fermionic and gauge structures
makes the
Standard Model look like an improbable fundamental theory.  To put it in
a less
qualitative way, the Standard Model contains many free parameters ({\it e.g.}
the
three gauge coupling constants, the nine fermion masses and the four
Cabibbo--Kobayashi--Maskawa mixing parameters); these correspond to
important
physical quantities, but cannot be computed in the context of the model.
Simplifying the Standard Model structure and predicting its free
parameters are
therefore basic tasks of a successful theory.

In these lectures I review the main current ideas about theories
beyond
the Standard Model, keeping the discussion at a qualitative level and
making no
use of advanced mathematics.  More comprehensive reviews can be found in
refs.
\cite{gut} (for GUTs), \cite{susy} (for supersymmetry) and \cite{tec}
(for
technicolour). 

\section{GUT $SU_5$}

The first attempts to extend the structure of the Standard Model have
led to
the construction of Grand Unified Theories (GUTs) \cite{gg}.  The basic
idea
is that gauge interactions are described by a single simple gauge group,
which
contains the Standard Model $SU_3 \times SU_2 \times U_1$ as a subgroup
and as
a low-energy manifestation.  At first this may seem impossible, since a
simple
gauge group contains a single coupling constant $g_X$ and the strong,
weak and
electromagnetic couplings have different numerical values.  However 
it should be
remembered that, in a quantum field theory, the coupling constants depend
on
the energy scale at which they are probed, as a consequence of the
exchange of
virtual particles surrounding the charge.  
The evolution of the gauge coupling constants as a function of the
energy scale can be computed using renormalization group techniques and 
perturbation theory, and the relevant equations are described in 
sect.~7. There, we will also find that, as
we include the quantum effects of all Standard Model particles, the
three gauge coupling constants approach one another as the energy scale 
is raised. For the moment, let us assume that the three gauge couplings
meet at a single value for a specific energy scale ($M_X$) and
study possible GUT candidates describing the physics
above $M_X$ with a single gauge coupling constant $g_X$.

The simplest example of a GUT is based on the group $SU_5$.  Each fermion
family is contained in a ${\bf 10}$ + ${\bf {\bar 5}}$ representation
of
$SU_5$.  This can be understood from the decomposition in terms of the
Standard
Model group:
\begin{equation}
\matrix{
SU_5 & \rightarrow & SU_3 \times SU_2 \times U_1 \cr
{\bf 10} &\rightarrow & ({\bf {\bar 3}, 1,-\frac{2}{3}})
_{u^c_R} + ({\bf 3,2,\frac{1}{6}})_{q_L} + ({\bf 1,1,1})_{e^c_R} \cr
{\bf {\bar 5}} &\rightarrow & ({\bf {\bar 3}, 1, \frac{1}{3}})_{d^c_R} +
({\bf
1,2,-\frac{1}{2}})_{\ell_L}~.}
\label{rappr}
\end{equation}
Here the numbers inside the brackets respectively denote the $SU_3$ and
$SU_2$
representations and the $U_1$ quantum numbers. Equation (\ref{rappr})
shows that the degrees of
freedom for
all the (left-handed) fields in one Standard Model family are described
by the two $SU_5$
fields ${\bf 10}$ and ${\bf {\bar 5}}$.  In GUTs not only is the gauge
group 
unified, going from $SU_3 \times SU_2 \times U_1$ to $SU_5$ in this
specific
example,  but also the fermionic spectrum is simplified.  As 
quarks in QCD come
with different colours, in GUTs different quarks and leptons are just
different
aspects of the same particle.
This also explains the simple integer
relations
among the electric charges of different quarks and leptons.

\section{EXPERIMENTAL TESTS FOR GUTs}

Theoretical elegance is of course not a sufficient argument to convince 
us that
GUTs have anything to do with Nature.  We need to establish GUTs
predictions which
can be confronted with experimental data.  The basic idea of GUTs, gauge
coupling
unification, provides such a prediction.  Indeed at the GUT scale $M_X$
we can compute
the weak mixing angle:
\begin{equation}
\sin^2\theta_W \equiv \frac{e^2}{g^2} = \frac{{\rm Tr}(T^2_3)}{{\rm
Tr}(Q^2)} =
\frac{3}{8}~.
\label{senow}
\end{equation}
Here $T_3$ is the third isospin-component and $Q$ is the electric
charge.  The
trace in eq.~(\ref{senow}), taken over any $SU_5$ representation, follows from a
correct
normalization of the GUT generators.  Before comparing eq.~(\ref{senow}) with
experiment,
one has to rescale it to the low energies where coupling constants are
measured.  We
will do this in sect.~7, and show that 
eq.~(\ref{senow}) gives a successful
prediction for a
class of theories which we have not yet introduced, supersymmetric GUTs.
We anticipate here that, if gauge coupling unification has any chance to
succeed, the
unification scale $M_X$ must be extremely large, of the order of
$10^{15}$--$10^{16}$ GeV,
which, in the thermal history of our Universe, brings us to consider
events
occurring in the first $10^{-35}$--$10^{-38}$ s.

Since we have promoted the gauge group to $SU_5$, we expect new gauge
bosons and
therefore new forces which may have experimental consequences.
The decomposition of the
$SU_5$ gauge bosons in terms of Standard Model ones is:
\begin{equation}
\matrix{
SU_5 & \rightarrow & SU_3 \times SU_2 \times U_1 \cr
{\bf 24} & \rightarrow & ({\bf 8,1,0})_g + ({\bf 1,3,0})_W + ({\bf
1,1,0})_B +
({\bf 3,2,-\frac{5}{6}})_X + ({\bf {\bar 3}, 2, \frac{5}{6}})_{\bar X}}~.
\end{equation}
Together with the familiar degrees of freedom for the gluons $(g)$ and
the
electroweak gauge bosons $(W^\pm,W^0,B)$, we find new particles ($X$ and
${\bar X}$) which carry both colour and weak quantum numbers. The gauge
bosons $X$ and $\bar X$
affect weak interactions, but modify standard processes only by an
amount
$(M_W/M_X)^2$, a fantastically small number, whose effect is 
completely undetectable even
in the
most precise measurements.  Nevertheless, the $X$-mediated interactions
may not be
so invisible.  Let us inspect the interactions between $X, {\bar X}$ and
the
fermionic currents, which are dictated by $SU_5$ gauge invariance:
\begin{eqnarray}
{\cal L} &= & \frac{g_X}{\sqrt 2} \left\{
X^\alpha_\mu \left[ \bar d_{R\alpha} \gamma^\mu e^c_R + \bar d_{L
\alpha}
\gamma^\mu e^c_L + \varepsilon_{\alpha\beta\gamma} \bar u^{c\gamma}_L
\gamma^\mu
u^\beta_L \right] \right.+ \nonumber \\
&+&  \left. \bar X^\alpha_\mu \left[ - \bar d_{R \alpha} \gamma^\mu
\nu^c_R - \bar
u_{L \alpha} \gamma^\mu e^c_L + \varepsilon_{\alpha\beta\gamma} \bar
u^{c \gamma}_L
\gamma^\mu d^\beta_L \right] + \rm {h.c.} \right\}~.
\label{xlag}
\end{eqnarray}
Notice that one cannot assign a conserved baryon $(B)$ and lepton $(L)$
quantum
number to $X$ and $\bar X$; the new interactions violate both $B$ and
$L$.  In the
Standard Model $B$ and $L$ are accidental global symmetries, in the
sense that they
are just a consequence of gauge invariance and renormalizability.
It is not surprising that $B$ and
$L$ are
then violated in extensions of the Standard Model, in particular in GUTs
where
quarks and leptons are different aspects of the same particle.

The experimental
discovery of processes that violate $B$ and $L$ would be clear
evidence for
physics beyond the Standard Model.  One of the most important of such
processes is
proton decay, which has the dramatic consequence that ordinary matter is
not
stable.  It is easy to see from eq.~(\ref{xlag}) that 
the $X$ boson mediates the
transition
$uu \rightarrow e^+{\bar d}$.  When dressed between physical hadronic states,
this
transition is converted into the proton decay modes $p \rightarrow
e^+\pi^0,
e^+\rho^0, e^+\eta, e^+\pi^+\pi^-$, and so on.  The calculation of the proton
lifetime
yields
\begin{equation}
\tau_p = (0.2 - 8.0) \times 10^{31} \left( \frac{M_X}{10^{15}~{\rm GeV}}
\right)^4~{\rm yr}~.
\label{pro}
\end{equation}
The uncertainties in the numerical coefficient in eq.~(\ref{pro})
come mainly 
from the difficulty in estimating the matrix elements relating quarks
to
hadrons.  For reasonable GUT masses, $M_X \simeq 10^{15}$--$10^{16}$ GeV, 
eq.~(\ref{pro})
predicts a proton lifetime $10^{21}$--$10^{25}$ times 
larger than the age of the Universe.
It is fascinating that experiments can probe such slow processes by
studying very
large samples of matter.  The present experimental bound on the lifetime
of the decay
mode $p \rightarrow e^+\pi^0$, the dominant proton decay channel in
$SU_5$, is
\cite{pdg}
\begin{equation}
\tau(p \rightarrow e^+\pi^0) > 5.5 \times 10^{32}~{\rm yr}~.
\end{equation}
This bound already sets important constraints on possible GUT models.

GUTs also provide a framework in which the creation of a primordial baryon
asymmetry can
be understood and computed.  Although this is not an experimental test, it
is clearly
a very attractive theoretical feature.
Observations tell us that the present ratio of baryons to photons in the
Universe
is a very small number, $n_B/n_\gamma = 4$--$7 \times 10^{-10}$.  If
$n_B/n_\gamma$ is
then extrapolated back in time following the thermal history of the
Universe, one
finds that the excess of baryons over antibaryons at the time of the big
bang must
have been $\Delta_B \equiv (n_B - n_{\bar B})/n_B \sim 3 \times
10^{-8}$.  We find it
disturbing to consider that the present observed Universe is determined
by a
peculiar initial condition prescribing that for each three hundred
million baryons
there are three hundred million minus one antibaryons.

The hypothesis of baryogenesis is that $\Delta_B = 0$ at the time of the
big bang
and that
the small cosmic baryon asymmetry was dynamically created during the
evolution of
the Universe.  The physics responsible for the creation of $\Delta_B$
must
necessarily involve interactions which violate $B$.  GUTs are therefore
a natural
framework for baryogenesis and it has been proved \cite{bar} that they
have all the
necessary ingredients to generate the observed value of the present
baryon
density.

\section{$SO_{10}$ AND NEUTRINO MASSES}

 I have presented $SU_5$ as the simplest GUT, but  models based on larger
groups can
also be constructed. Probably the most interesting of them \cite{sod} 
is based
on the
orthogonal group $SO_{10}$, which contains $SU_5$ as a subgroup.  The
16-dimensional spinorial representation of $SO_{10}$ decomposes into
${\bf 10} +
{\bf{\bar 5}} + {\bf 1}$ under $SU_5$.  We recognize the fermion content
of one
Standard Model family.  It is quite satisfactory that quarks and leptons
with their
different quantum number assignments can be described by a single
$SO_{10}$
particle, for each generation.  

In addition to the ordinary quarks and leptons contained in the ${\bf 10
+ {\bar
5}}$ of $SU_5$, the spinorial representation of $SO_{10}$ contains also
a gauge
singlet.  This can be interpreted as the right-handed component of the
neutrino, allowing the possibility of Dirac neutrino masses.
The neutrino mass term can now be written in the form
\begin{equation}
({\bar \nu}_L {\bar \nu}^c_L ) {\cal M} \matrix{
\left( \nu^c_R \\ \nu_R \right)} + {\rm h.c.}~,
\end{equation}
where, for simplicity, we are considering only the one-generation case.
The
different entries of the neutrino mass matrix $\cal M$
\begin{equation}
{\cal M} =\left( \matrix{
 T & D \cr
D^T & S }\right)
\label{neu}
\end{equation}
can be understood in terms of symmetry principles.  The term $S$
transforms as a
singlet under the Standard Model gauge group and therefore is naturally
generated at the scale where the $SO_{10}$ symmetry is broken, $S \sim
M_X$.  The
other two terms, $T$ and $D$, transform respectively as a triplet and a
doublet under
the weak group $SU_{2}$; therefore they can be generated only after the
Standard
Model gauge group is broken.  However, vacuum expectation values of
triplet fields
lead to an incorrect relation between the strengths of
neutral and charged weak currents.
We
conclude therefore that $T \simeq 0$ and $D \simeq m_f$, where $m_f$ is 
a typical
fermion (quark or charged lepton)
mass.  After diagonalization of the matrix in eq.~(\ref{neu}), we find
one heavy
eigenstate with mass of order $M_X$ and one (mainly left-handed)
eigenstate with mass \cite{see}:
\begin{equation}
m_\nu \simeq \frac{m^2_f}{M_X} = 10^{-6}{\rm eV}
\left( \frac{m_f}{\rm GeV} \right)^2
\left( \frac{10^{15}\,{\rm GeV}}{M_X} \right) ~.
\end{equation}

In the context of the $SO_{10}$ GUT, not only do we expect neutrinos to be
massive, but we
also understand in terms of symmetries why their masses must be much
smaller than
the typical scale of the other fermion masses.

\section{THE HIERARCHY PROBLEM}

The hierarchy (or naturalness) problem \cite{natural} 
is considered to be one of the most
serious
theoretical drawbacks of the Standard Model and most of the attempts to
build
theories beyond the Standard Model have concentrated on its solution.
It springs
from the difficulty in field theory in keeping fundamental scalar particles
much
lighter than $\Lambda_{\rm max}$, the maximum energy 
scale up to which the theory remains valid.

It is intuitive to require that if a
particle mass is
much smaller than $\Lambda_{\rm max}$, there should exist a (possibly
approximate)
symmetry under which the mass term is forbidden. We know an example of
such a
symmetry for spin-one particles.  The photon is, theoretically speaking,
naturally
massless since the gauge symmetry $A_\mu \rightarrow A_\mu +
\partial_\mu\lambda$
forbids the occurrence of the photon mass term $m^2A_\mu A^\mu$.
Similarly, we can
identify a symmetry which protects the mass of a fermionic particle.  A
chiral
symmetry, under which the left-handed and right-handed fermionic
components
transform differently $\psi_L \rightarrow e^{i\alpha} \psi_L, \psi_R
\rightarrow
e^{i\beta} \psi_R, \alpha \ne \beta$, forbids the mass term $m\, \bar
\psi_L \psi_R$
+ h.c. Scalar particles can be naturally light if they are Goldstone
bosons of
some broken global symmetry since their non-linear transformation
property
$\varphi \rightarrow \varphi + a$ forbids the mass term $m^2 \varphi^2$.

In the case of the Higgs particle, required in the Standard Model by the
electroweak symmetry
breaking mechanism,
we cannot rely on any of the above-mentioned
symmetries.  In
the absence of any symmetry principle, we expect the Higgs potential 
mass parameter $m^2_H$
to be of the order of $\Lambda^2_{\rm max}$.  Even if we
artificially
set the classical value of $m^2_H$ to zero, it will be generated by
quadratically
divergent quantum corrections:
\begin{equation}
m^2_H = \frac{\alpha}{\pi}~\Lambda^2_{\rm max}~,
\label{ft}
\end{equation}
where $\alpha$ measures the effect of a typical coupling constant.

One may argue that in a renormalizable theory, the bare value of
any parameter
is an infinite (or, in other words, cut-off dependent) quantity, 
without a precise
physical
meaning.  Since all divergences can be reabsorbed, one can just choose
the
renormalized quantity to be equal to any appropriate value.  However, we
believe
that a complete description of particle interactions in a final theory
will be free
from divergences.  From this point of view, the cancellation between
a bare
value and quadratically divergent quantum corrections looks like a
conspiracy
between the infra-red (below $\Lambda_{\rm max}$) and the ultraviolet
(above
$\Lambda_{\rm max}$) components of the theory.  We do not accept such a
conspiracy,
but, on the other hand, we know that the parameter $m^2_H$ sets the
scale for
electroweak symmetry breaking and it is therefore directly related to
$m^2_W$.
We thus require that the quantum corrections in eq.~(\ref{ft}) do
not exceed
$m^2_W$.  This implies an upper bound on $\Lambda_{\rm max}$:
\begin{equation}
\Lambda_{\rm max} \lappeq \sqrt{ \frac{\pi}{\alpha}} M_W \simeq {\rm
TeV}~.
\label{te}
\end{equation}
We can conclude that the Standard Model has a natural upper bound
at the TeV
scale, where new physics should appear and modify the ultraviolet
behaviour of the
theory.

The hierarchy problem becomes most apparent when one considers GUTs.
Here the
Higgs potential of the model contains two different mass parameters:
one is of
order $M_X$ and sets the scale for the breaking of the unified group; the
other is
of order $M_W$ and sets the scale for the ordinary electroweak breaking.
By
explicit calculation, one can show \cite{gild} that these parameters mix
at the
quantum level and the hierarchy of the two mass scales can be maintained
only at
the price of fine-tuning the parameters by an amount $(M_X/M_W)^2$.

\section{SUPERSYMMETRY}

Supersymmetry \cite{sup}, contrary to all other ordinary symmetries in field
theory,
transforms bosons to fermions and vice versa.  This means that bosons and
fermions
sit in the same supersymmetric multiplet.  In the simplest version of
supersymmetry
(the so-called $N = 1$ supersymmetry), each complex scalar has a Weyl fermion
companion and
each massless gauge boson also has a Weyl fermion companion;  similarly
the spin-2
graviton has a spin-3/2 companion, the gravitino.  Invariance under
supersymmetry
implies that particles inside a supermultiplet are degenerate in mass.
It is
therefore evident that, in a supersymmetric theory, if a chiral symmetry
forbids
a fermion mass term, it forbids also the appearance of a scalar mass term,
such as the
notorious Higgs mass parameter. The hierarchy problem discussed
in the
previous section can now be solved.  Indeed, it has been 
proved that a supersymmetric
theory is free
from quadratic divergences \cite{quad}.  The contribution to 
$m^2_H$ proportional to
$\Lambda^2_{\rm max}$ in eq.~(\ref{ft}) coming from a bosonic loop is
exactly
cancelled by a loop involving fermionic particles.  Since the dependence
on
$\Lambda^2_{\rm max}$ has now disappeared, we can extend the scale of
validity of
the theory 
without provoking any hierarchy problem.

It should also be mentioned that when supersymmetry is promoted to a
local
symmetry, which means that the transformation parameter depends on
space-time, then
the theory automatically includes gravity and is called supergravity.
Because of this characteristic,
supersymmetry is believed to be a necessary ingredient for the
complete
unification of forces.

Here we are interested in the minimal extension of the Standard Model
compatible
with supersymmetry.  Each Standard Model particle is accompanied by a
supersymmetric partner:  scalar particles (squarks and sleptons) are the
partners
of quarks and leptons, and fermion particles ({\it e.g.} gluinos) are the
partners of the Standard Model bosons ({\it e.g.} gluons).
Supersymmetry also
requires two Higgs doublets, as opposed to the single Higgs doublet of
the
Standard Model, and their fermionic partners mix with the fermionic
partners of the
electroweak gauge bosons to produce particles with one unit of
electric charge (charginos) or no electric charge (neutralinos).

Supersymmetry ensures that the couplings of all these new particles are
strictly
related to ordinary couplings.  For instance, the couplings of
squarks to one
or two gluons, of gluinos to gluons, of squarks and gluinos to quarks
are solely
determined by $\alpha_s$, the QCD gauge coupling constant.

The supersymmetric generalization of the Standard Model is therefore a
well-defined theory where all new interactions are described by the
mathematical
properties of the supersymmetric transformation.  As such, however, 
the theory is not
acceptable since it predicts a mass degeneracy between the ordinary and
the supersymmetric
particles;  in Nature, therefore, supersymmetry is 
not an exact symmetry.  In order
to
preserve the solution of the hierarchy problem we need to break
supersymmetry while maintaining the
good ultraviolet behaviour of the theory.  It has been shown
\cite{gir} that if only a certain set of supersymmetry-breaking terms
with
dimensionful couplings are introduced, then the quadratic divergences still
cancel, but the mass degeneracy is removed.
Let us generically call $m_S$ the mass that sets the scale for the
dimensionful
couplings which softly break supersymmetry.  This scale has a definite
physical
meaning, since all new supersymmetric particles acquire masses of order
$m_S$.
It is the energy scale at which supersymmetry has to be looked for in
experiments.

By explicit calculation one finds that, in a softly broken
supersymmetric theory,
quadratic divergences cancel, but some finite terms of the
kind
$(\alpha/\pi)m^2_S$ remain.
From
eq.~(\ref{ft}) we recognize that $m_S$ behaves as the cut-off of
quadratic divergences
in the Standard Model.  This is not entirely surprising since, in the
limit $m_S
\rightarrow \infty$, all supersymmetric particles decouple and one
should recover the
ultraviolet behaviour of the Standard Model.  Therefore we conclude
that, in a softly broken supersymmetric
theory,  the
cut-off of quadratic divergences has a
physical meaning since it is related to $m_S$, the mass scale of the new
particles.
Moreover, following the same argument that led us to eq.~(\ref{te}) we
find that
these new particles cannot be much heavier than the TeV scale, if
supersymmetry
solves the hierarchy problem.  In sect.~8, I will make this argument
more
quantitative.

Although technically successful, it may appear 
that the
introduction of the soft supersymmetry-breaking terms is too arbitrary
to be entirely
satisfactory. But, on the contrary, it has 
a very appealing explanation \cite{bn}.  Let us first 
promote supersymmetry to supergravity, 
possibly a
necessary step towards complete unification of forces.  Then assume
that
supergravity is either spontaneously or dynamically broken in a sector
of the theory
that does not directly couple to ordinary particles.  In this case,
gravity
communicates the supersymmetry breaking, and the low-energy effective
theory of the
supersymmetric Standard Model contains exactly all the terms which
break
supersymmetry without introducing quadratic divergences.

From this point of view, the appearance of the soft-breaking terms can
be understood
in terms of well-defined dynamics.  However, we do not yet know which
mechanism 
breaks supersymmetry and therefore we are not able to compute the
soft-breaking
terms.  This is unfortunate because these define the mass spectrum of the
new
particles.  All we can do now is to keep them as free parameters and
hope they will
be determined by experimental measurements or calculated, if theoretical
progress is
made.  In the minimal version of the theory, there are only four such
parameters but,
if some assumptions are relaxed, the number of free parameters can grow
enormously.

\section{SUPERSYMMETRIC UNIFICATION}

In the previous section, we have extended the Standard Model to include
supersymmetry
in order to solve the hierarchy problem.  We can now incorporate 
within this model the
ideas of grand unification, and construct a supersymmetric GUT \cite{susygut}.  

As discussed in
sect.~3, the first test of a GUT is gauge coupling unification.  At
the one-loop
approximation the evolution of the $SU_3 \times SU_2 \times U_1$ gauge
coupling
constants with the energy scale $Q^2$ is given by
\begin{equation}
\frac{d \alpha_i}{dt} = - \frac{b_i}{4\pi}~\alpha^2_i ~~~~~~\Rightarrow
~~~~~~\alpha_i(t) =
\frac{\alpha_i(0)}{ 1 + \frac{b_i}{4\pi} \alpha_i(0)t}~,~~~~~~i = 1,2,3,
\end{equation}
where $t = \log (M^2_X/Q^2)$.  The coefficients $b_i$ take into account
the numbers
of degrees of freedom and the gauge quantum numbers of all particles
involved in
virtual exchanges.  For the Standard Model, we find
\begin{equation}
b_3 = -7 + \frac{4}{3} (N_g-3)~, \;\;\;b_2 = - \frac{19}{6} + \frac{4}{3}
(N_g - 3)~,
\;\;\; b_1 = \frac{41}{6} + \frac{20}{9} (N_g - 3)~,
\end{equation}
where $N_g$ is the number of generations.  In the supersymmetric case
all new
particles influence the running of the gauge coupling constants and
modify the
$b_i$ parameters,
\begin{equation}
b_3 = -3 + 2(N_g -3)~, \;\;\; b_2 = 1 + 2(N_g-3)~, \;\;\; b_1 = 11 +
\frac{10}{3} (N_g-3)~.
\end{equation}
Assuming $N_g = 3$ and gauge coupling unification, {\it i.e.} $\alpha_3(0) =
\alpha_2(0) = 5/3
\alpha_1(0)$, we can compute the QCD coupling $\alpha_s(M_Z) (\equiv
\alpha_3(M_Z))$ and
$\sin^2\theta_W (\equiv [1 + \alpha_2(M_Z)/$$\alpha_1(M_Z)]^{-1})$ as a
function of
$M_X$, taking $\alpha^{-1}(M_Z) = 127.9 \pm 0.1~(\alpha^{-1} \equiv
\alpha^{-1}_1 +
\alpha^{-1}_2)$.  The results of the theoretical calculations in the
Standard and
supersymmetric models are shown in fig.~1, together with the experimental
result \cite{pdg}.
Unification of couplings is clearly inconsistent with the Standard Model
evolution
for any value of $M_X$.  This rules out any simple GUT which breaks directly
into $SU_3
\times SU_2 \times U_1$, with only ordinary matter content.  Inclusion of
additional light
particles or intermediate steps of gauge symmetry breaking may
reconcile the
Standard Model with the idea of unification.  Of course, in this case,
any prediction
from gauge coupling unification is necessarily lost.  More interesting
is the
supersymmetric case in which unification is achieved in the minimal
version of the
model, with $M_X \simeq 10^{16}$~GeV.  From the historical point of view,
it is amusing to notice
that in 1981, when 
supersymmetric GUTs were first proposed, the experimental data \cite{data}
were compatible
with standard GUTs, but disfavoured supersymmetric unification; see
fig.~1.

\section {ELECTROWEAK SYMMETRY BREAKING}

As a realistic theory of particle interactions, the supersymmetric model
should
describe the correct pattern of electroweak symmetry breaking.  This is
obtained by
the Higgs mechanism.  As already mentioned in sect.~6, supersymmetry
requires two
Higgs doublets, as opposed to the single one of the Standard Model.
Along the
neutral components of the two Higgs fields, the scalar potential is:
\begin{equation}
V(H^0_1,H^0_2) =
m^2_1 |H^0_1|^2 + m^2_2 |H^0_2|^2 - m^2_3 (H^0_1H^0_2 + {\rm h.c.}) +
\frac{g^2+g^{\prime}}{8} \left( |H^0_1|^2 - |H^0_2|^2 \right)^2
\label{P}
\end{equation}
where $g, g^\prime$ are respectively the $SU_2$ and $U_1$ gauge coupling
constants.  The
mass parameters $m^2_1, m^2_2$ and $m^2_3$ originate from soft-breaking
terms and are
therefore of the order of $m_S$, the mass scale introduced in 
sect.~6.
The stability of the potential for large values of fields along the
direction $H^0_1
= H^0_2$ requires
\begin{equation}
m^2_1 + m^2_2 > 2|m^2_3|~.
\label{A}
\end{equation}
Since electroweak symmetry is broken, the origin $H^0_1 = H^0_2 = 0$
must correspond
to an unstable configuration, which implies:
\begin{equation}
m^2_1 m^2_2 < m^4_3~.
\label{B}
\end{equation}

It is often assumed that the soft-breaking terms satisfy some
universality conditions
around $M_X$.  Notice that, should for instance $m^2_1 = m^2_2$,
eqs.~(\ref{A}) and
(\ref{B}) cannot be simultaneously satisfied and electroweak symmetry
remains
unbroken.  Nevertheless, before drawing any conclusion, we have to
include the
renormalization effects of changing the scale from $M_X$ to the
electroweak scale
$M_W$.  These effects are important as they are proportional to a large
logarithm,
$\log (M^2_X/M^2_W)$, and they have been systematically computed up to
two loops
\cite{two}.  Generically, the effect of gauge interactions is to
increase the masses
as we evolve from $M_X$ to $M_W$.  Therefore, if all masses are equal at
$M_X$, we
expect gluinos to be heavier than charginos and neutralinos, and
similarly squarks to
be heavier than sleptons, because of the dominant QCD effects.  On the
other hand,
Yukawa interactions decrease the masses in the renormalization from high
to low
energies.  Therefore, the stops will be the lightest among squarks,
since the top
quark coupling gives the dominant Yukawa effect.  

Let us now consider
the
evolution of the Higgs mass parameters.  As they do not feel QCD forces
at one
loop, their gauge renormalization is not very significant.  The Yukawa coupling
effect is
important for $m^2_2$, because $H_2$ is the Higgs field responsible for
the top
quark mass, but not for $m^2_1$.  Therefore, as an effect of the heavy
top quark,
$m^2_2$ decreases and it is likely to be driven negative around
the weak
scale, while $m^2_1$ remains positive.  For $m^2_1 > 0$ and $m^2_2 < 0$,
eqs.~(\ref{A}) and (\ref{B}) can be easily satisfied and electroweak
symmetry is
broken \cite{ross}.

In conclusion, the supersymmetric model is consistent with electroweak
symmetry
breaking and the mechanism involved is appealing in several ways.  First
of all,
the breaking is driven by purely quantum effects, a theoretically
attractive
feature.  Then it needs a heavy top quark, which agrees with the Tevatron
discovery.  Finally, we have found that the dynamics itself chooses to
break down
$SU_2$.  In a supersymmetric theory, colour $SU_3$ could spontaneously
break if
squarks get a vacuum expectation value, but this does not happen 
since
squark masses squared receive large positive radiative corrections.

The minimization of the Higgs potential in eq.~(\ref{P}) gives:
\begin{equation}
\frac{M^2_Z}{2} \equiv \frac{g^2+g^{\prime}}{8}~v^2 =
\frac{m^2_1 - m^2_2 \tan^2\beta}{\tan^2\beta - 1}~,
\label{C}
\end{equation}

\begin{equation}
\sin^2\beta = \frac{2m^2_3}{m^2_1 + m^2_2}~,
\label{D}
\end{equation}
where
\begin{equation}
\langle H_1^0 \rangle = \frac{v}{\sqrt 2} \cos \beta~,~~\langle H_2^0
\rangle =
\frac{v}{\sqrt 2} \sin\beta ~.
\end{equation}
Equation~(\ref{C}) can be interpreted as a prediction of $M_Z$ in terms
of the
soft supersymmetry-breaking parameters $(a_i)$ which determine $m^2_1,
m^2_1$,
and $m^2_3$.  Unfortunately, we are not able to compute supersymmetry
breaking,
and therefore we can only use eq.~(\ref{C}) as a constraint which fixes
one of
the parameters $a_i$ in terms of the others.

We can also use eq.~(\ref{C}) to define a quantitative criterion 
for obtaining upper bounds on supersymmetric particle masses from the
naturalness requirement
\cite{bg}.  It is intuitive that, as
the supersymmetry-breaking scale $m_S$
grows,
eq.~(\ref{C}) can hold only with an increasingly precise cancellation
among the
different terms.  We therefore require, for each parameter $a_i$:
\begin{equation}
\left | \frac{a_i}{M^2_Z}~\frac{\partial M^2_Z}{\partial a_i} \right| <
\Delta ~,
\label{E}
\end{equation}
where $M^2_Z$ is given by eq.~(\ref{C}) and $\Delta$ is the degree of
fine
tuning.  Equation~(\ref{E}) can now be translated into upper bounds on the
supersymmetric particle masses.  Independently of specific universality
assumptions on supersymmetry-breaking terms, we find \cite{dg}, for instance,
that the
chargino and the gluino are respectively lighter than 120 and 500~GeV,
if fine
tunings no greater than 10\% $(\Delta = 10)$ are required.

\section{HIGGS SECTOR}

Supersymmetry requires two Higgs doublets and therefore an extended
spectrum of
physical Higgs particles.  Out of the eight degrees of freedom of the
two complex
doublets, three are eaten in the Higgs mechanism and five correspond to
physical
particles.  These form two real CP-even scalars $(h,H)$, one
real CP-odd
scalar $(A)$, and one complex scalar $(H^+)$.  As we have seen in the
previous
section, the Higgs potential contains three parameters $(m^2_1, m^2_2,
m^2_3)$
and one of them is fixed by the electroweak symmetry-breaking condition,
eq.~(\ref{C}).  Therefore, all tree-level masses and gauge
couplings of the five Higgs
particles
are completely described by only two free parameters.

Another important feature of the supersymmetric Higgs potential is that
the
quartic coupling is given in terms of gauge couplings, see
eq.~(\ref{P}).  In the
Standard Model case, the quartic Higgs coupling measures the Higgs mass.
Therefore, it is not surprising to find that in supersymmetry the mass
of the
lightest Higgs is bounded from above:
\begin{equation}
m_h < M_Z |\cos 2\beta|~.
\label{HIG}
\end{equation}
Supersymmetry does not only provide a solution to the hierarchy problem
by stabilizing the Higgs mass parameter, but also predicts the existence of
a Higgs
boson lighter than the $Z^0$.

Note that eq.~(\ref{HIG}) holds only 
at the classical level.  There are important
radiative corrections to the lightest Higgs mass
proportional to $m^4_t$ \cite{higgs}:
\begin{equation}
\delta m^2_h \simeq \frac{3}{\pi^2}~\frac{m^4_t}{v^2}~\log
\frac{m_S}{v}~.
\end{equation}
The upper bound given in eq.~(\ref{HIG}) is then modified, and the
result is
shown in fig.~2 \cite{quiros}.  
For extreme values of the parameters, $m_h$ can be as
heavy as
150~GeV, but it is generally much lighter.

This is an excellent opportunity for LEP2, where the
Standard
Model Higgs boson can be discovered 
via the process $e^+e^- \rightarrow h Z^0$ in
essentially the entire
kinematical range $m_h < \sqrt s - M_Z$.
In the supersymmetric case, the search is more involved, because of
the extended Higgs
sector.
For $\tan \beta$
close to
1, the supersymmetric Higgs boson resembles the Standard Model
counterpart and
the LEP2 search is unchanged.  For large values of $\tan \beta$, the
cross-section for $e^+e^- \rightarrow h Z^0$ is reduced and can become
unobservable at
LEP2.  However, at the same time, the CP-odd Higgs boson $A$ becomes
light and the
cross-section for the process $e^+e^- \rightarrow hA$ is then sizeable.  The
two different
Higgs production mechanisms are therefore complementary and allow the
search for the
supersymmetric Higgs boson at LEP2 for most of the parameters.  Nevertheless,
a complete
exploration of the whole supersymmetric parameter space will be possible
only at the LHC, at
the beginning of the next millenium.

The discovery of a light Higgs boson is certainly not a proof of the
existence of supersymmetry at low energies.
However, in the Standard Model, vacuum stability 
imposes a {\it lower} bound
on the Higgs
mass as a function of the top quark mass \cite{cab}.  This is shown in fig.~2 
\cite{quiros},
where the
validity of the Standard Model is assumed 
up to the Planck mass.  For comparison, the
{\it upper} bound
on the supersymmetric Higgs mass is also shown in fig.~2.  
Notice that, for $m_t < 175$~GeV, the Higgs discovery can discriminate
between the supersymmetric model and the Standard Model with $\Lambda_{\rm
max}=M_{Pl}$.
Although this is
not strictly true for $m_t > 175$~GeV, it is clear that the Higgs search
can in general give good
indications about the scale of new physics.

\section{SUPERSYMMETRY AND EXPERIMENTS}

If the Higgs search is certainly an important experimental test, evidence
for
low-energy supersymmetry will come only from the discovery of the
partners of
ordinary particles.

The most important feature of supersymmetry phenomenology is the
existence of a
discrete symmetry, called $R$-parity, which distinguishes ordinary
particles from
their partners.  This is not an accidental symmetry, in the sense that it
is not an
automatic consequence of supersymmetry and gauge invariance.
Nevertheless, it is
usually assumed, or else dangerous $B$- or $L$-violating interactions are
introduced.
It can be understood as a consequence of gauge symmetry in GUT models
which contain
left-right symmetric groups.  If $R$-parity is indeed conserved only an
even number
of supersymmetric partners can appear in each interaction.  As a
consequence,
supersymmetric particles are produced in pairs and the lightest
supersymmetric
particle is stable.

In most of the models, this stable particle turns out to be the lightest
neutralino
$(\chi^0)$.  This is fortunate for the model, since the present 
density of electric- or
colour-charged
heavy particles is very strongly limited by searches for exotic atoms 
\cite{ellis}.
A stable
neutral particle is not only allowed by present searches but also
welcome since it can
explain the presence of dark matter in the Universe (see ref.~\cite{spiro}). 
From
the point of view of collider experiments, $\chi^0$ will behave as a
heavy neutrino
which escapes the detector, leaving an unbalanced momentum and missing
energy in
the observed
event.  The distinguishing signature of supersymmetry is therefore an excess
of missing
energy and momentum.  For example, in $e^+e^-$ colliders, charginos and
sleptons are
pair-produced with typical electroweak cross-sections and then decay,
giving rise to
events such as:
\begin{eqnarray}
e^+e^- \rightarrow \chi^+\chi^- 
&\rightarrow& {\rm isolated~leptons~and/or~ jets} +
E\llap{$/$}~,
\nonumber \\
e^+e^- \rightarrow \tilde \ell^+ \tilde \ell^- &\rightarrow & {\rm
isolated~leptons} +
E\llap{$/$}~.
\end{eqnarray}
Using these processes, LEP1, working at the $Z^0$ peak, was able to rule
out the
existence of these particles with masses less than $M_Z/2$ \cite{pdg}.  
LEP2 should
cover most of
the kinematical range, and discover or exclude $\chi^+$ and $\tilde \ell^+$
with masses almost up to $\sqrt s /2$.  This is certainly going to
be a very
critical region since, as we have seen in sect.~8, the 10\% 
fine-tuning limits place the weakly-interacting supersymmetric particles
at the border of the LEP2 discovery reach.

Strongly-interacting particles, such as squarks and gluinos, can be best
studied at
hadron colliders where they are produced with large cross-sections.  The
signature is
again missing transverse energy carried by the neutralinos produced in
the decays of
squarks and gluinos.  Tevatron experiments have set limits on the masses
of these
particles of about
150--200~GeV, depending on the particular model assumptions.  At the LHC
squarks and gluinos 
can be searched even for masses of several TeV, well above the
10\% fine-tuning limits.

It is worth pointing out that although $e^+e^-$ colliders are the ideal
machines for
a systematic search of new weakly-interacting particles, charginos and
neutralinos
may also be discovered at hadron colliders, for instance in the process:
\begin{equation}
p \bar p \rightarrow \chi^\pm_1 \chi^0_2~,~~~\chi^\pm_1 \rightarrow
\ell^\pm \nu
\chi^0_1~, ~~~ \chi^0_2 \rightarrow \ell^+\ell^- \chi^0_1 ~.
\end{equation}
The signal  of three leptons and missing transverse energy in the final
state has
almost no Standard Model background, when sufficient lepton isolation
requirements
are imposed.  However, it is difficult to obtain lower bounds on the new
particle
masses, because the leptonic branching ratios of charginos and
neutralinos depend strongly
on the model parameters.

In conclusion, this generation of colliders is testing the theoretically
best-motivated region of parameters in the supersymmetric model.  We can be
confident
that, after the LHC has run, either low-energy supersymmetry will have  been
discovered or it
must be discarded, since its main motivation is no longer valid.

\section{THE FLAVOUR PROBLEM}

The Standard Model Lagrangian for gauge interactions is invariant under
a global $U_3^5$ symmetry, with each $U_1$ acting on the generation
indices of the five irreducible fermionic representations of the gauge 
group $(q_L,u_R^c,d_R^c,\ell_L,e_R^c)_i$, $i=1,~2,~3$. 
This symmetry, called flavour (or generation) symmetry, implies that
gauge interactions do not distinguish among
the three generations of quarks and leptons.
In the real world, this symmetry must be broken, as 
quarks and leptons of different generations have different masses. 
However, the breaking must be such as to maintain an approximate
cancellation of Flavour-Changing Neutral Currents (FCNC). This is
called the flavour problem.

In the Standard Model the flavour problem is solved in a simple and
rather elegant way. The flavour symmetry is broken only by the Yukawa
interactions between the Higgs field and the fermions. After electroweak
symmetry breaking, these interactions give rise to the various masses
of the three generations of quarks and leptons. The attractive feature
of this mechanism is that all FCNC exactly
vanish at tree level \cite{gim}. This is a specific 
property of the Standard Model
with minimal Higgs structure and it is not automatic in models with an
enlarged Higgs sector. Small contributions to FCNC are generated at
loop level and generally agree with experimental observations. Athough
this mechanism provides a great success of the Standard Model, it prevents us
from computing any of the quark or lepton masses, as these are introduced in
terms of some free parameters.

In supersymmetry, the solution of the flavour problem is more arduous.
Most of the soft-breaking terms introduced in sect.~6 generally violate
the flavour symmetry and give too large contributions to the FCNC. 
This can be understood by recalling that, in a
softly-broken supersymmetric theory, the mass matrices for quarks and
squarks are independent and therefore cannot be simultaneously diagonalized by
an equal rotation of the quark and squark fields. Thus neutral currents
involving gluino--quark--squark vertices can mediate significant transitions
among the different generations. Only if squarks and gluinos were heavier
than 10--100 TeV could generic soft-breaking terms be consistent with
observations of FCNC processes. Since, as discussed in sect.~8, the very
motivation for low-energy supersymmetry implies that squarks and gluinos must 
be lighter than 500--1000 GeV, we have to postulate that the 
supersymmetry-breaking terms have some specific property.

The first possibility is that the supersymmetry-breaking terms respect
the flavour symmetry in the limit of vanishing Yukawa couplings. This
possibility is often advocated in models based on supergravity,
on the basis of the hypothesis that all gravitationally-induced interactions
are flavour-invariant. However, this hypothesis has been shown to be incorrect
both in supergravity models with generic K\"ahler metrics \cite{wein} and
in models derived from superstrings \cite{string}. Nevertheless, this is
an interesting possibility, since it significantly reduces the number of
free parameters in the supersymmetry-breaking terms and allows sharp
predictions testable at future colliders.

The other possibility is that the supersymmetry-breaking terms violate
the flavour symmetry but are approximately aligned with the corresponding
flavour violation in the fermionic sector ({\it e.g.} with the Yukawa
couplings). This can be the result of some new symmetry \cite{nir} or
some dynamical mechanism \cite{dgt}.

It is likely that the solution of the flavour problem is linked with
the mechanism of supersymmetry breaking and therefore it will only be
unravelled after significant theoretical developments. Now we can only
speculate that an understanding of the flavour problem may help us
to calculate the amount of flavour breaking and ultimately all quark
and lepton masses.

\section{TECHNICOLOUR}

We have seen how supersymmetry can cure the hierachy problem of the
Standard Model by
stabilizing the mass scale in the Higgs potential.  Technicolour \cite{tech} 
offers
a different
solution to the hierarchy problem, based on the idea of removing all
fundamental
scalar particles from the theory.  The mass scale which sets the
electroweak breaking
is dynamically determined in a strongly interacting gauge theory with
purely
fermionic matter.

The presence of light scalars (mesons) in the hadronic spectrum does
not pose a
problem of hierarchy.  The description of mesons as fundamental
particles is valid
only up to about $\Lambda_{\rm QCD}$.  Above this scale, physics is
described in terms of
quarks and gluons, and hadrons have to be interpreted as composite
particles.
Technicolour aims to describe the Higgs boson as a composite particle,
similarly to
the case of mesons in QCD.

In order to illustrate the main idea of technicolour, let us consider as
a toy model
QCD with only two massless flavours $(m_u = m_d = 0)$.  In this limit,
the theory has
a chiral $SU(2)_L \times SU(2)_R$ invariance, in which the left-handed and
right-handed
components of the up and down quarks are rotated independently.  As QCD
becomes
strongly-interacting at $Q^2 \lappeq \Lambda^2_{\rm QCD}$, the quark
condensates are formed:
\begin{equation}
\langle u \bar u \rangle = \langle d \bar d \rangle = {\cal 
O}(\Lambda^3_{\rm QCD})~.
\end{equation}
If the two condensates are equal, the chiral symmetry is broken to the
vectorial part
$SU(2)_{L+R}$.  Goldstone's theorem ensures the existence of three
massless scalar
particles in the spectrum, the pions $\pi^0, \pi^\pm$.  In the real
world, quark
masses explicitly break chiral symmetry and give small masses to the
pions.  Also, if
the strange quark is included, the chiral symmetry $SU(3)_L \times
SU(3)_R$ is broken
to $SU(3)_{L+R}$, giving rise to the meson octet as approximate Goldstone
bosons.

Let us turn on weak interactions in our toy model.  Since the $W$ boson
couples to
quarks, it also interacts with the pions.  This coupling can be obtained
from PCAC,
which determines the matrix element of the broken current $(j^a_\mu)$ in
terms of the
pion decay constant $f_\pi$:
\begin{equation}
\langle 0 | j^a_\mu|\pi^b \rangle = f_\pi q_\mu \delta^{ab}~.
\label{MAT}
\end{equation}
Here $a, b$ are $SU(2)$ indices and $q_\mu$ is the pion four-momentum.
From
eq.~(\ref{MAT}) and the coupling of the $W$ boson to the weak current,
we obtain the
coupling between $W^a_\mu$ and $\pi^b$:
\begin{equation}
\frac{g}{2} f_\pi q_\mu \delta^{ab}~.
\label{K}
\end{equation}
Consider now the correction of one-pion exchange in the $W$ propagator:
\begin{equation}
\frac{1}{q^2} + \frac{1}{q^2} \left( \frac{g}{2} f_\pi q^\mu \right)
\frac{1}{q^2}
\left( \frac{g}{2} f_\pi q_\mu \right) \frac{1}{q^2}~.
\label{L}
\end{equation}
The first term corresponds to the uncorrected massless $W$ propagator
and the second
term corresponds to the exchange of a massless pion between two $W$
propagators with
the coupling given in eq.~(\ref{K}).  We can insert an infinite number
of pion
exchanges, but it is not difficult to sum the whole series:
\begin{equation}
\frac{1}{q^2} \sum^\infty_{n=0} \left[ \left(
\frac{g}{2} f^2_\pi \right) \frac{1}{q^2} \right]^n = \frac
{1}{q^2 - \left(\frac{g}{2} f_\pi \right)^2}~.
\label{M}
\end{equation}
Equation~(\ref{M}) shows that the effect of the pion exchange is to
shift the pole
value of the $W$ propagator to
\begin{equation}
M_W = \frac{g}{2} f_\pi ~.
\label{N}
\end{equation}
The $W$ boson has acquired mass, which is not a surprising result if we
think that we
have promoted a global broken symmetry to a local invariance.  The value
for the $W$
mass given by eq.~(\ref{N}) is about 30~MeV, certainly too small to
explain the
experimental data.

We can use the result of this toy model and explain the physical value
of $M_W$, if we
introduce a new force, called technicolour. Technicolour behaves in a similar
fashion to the
ordinary colour forces but it becomes strong at a much larger scale
$\Lambda_{TC}
\simeq 500$~GeV.  The simplest technicolour model is very easy to
construct.  Take a
doublet of fermions with the same electroweak quantum numbers as the up
and down
quarks, assign to them a technicolour charge and call them techniquarks
$U$ and $D$.
The condensates
\begin{equation}
\langle \bar U U \rangle = \langle \bar D D \rangle = {\cal O}(\Lambda^3_{TC})
\end{equation}
generate three composite Goldstone modes, which become the longitudinal
degrees of
freedom of the $W$ and $Z$ gauge bosons.  We have then built a model of
electroweak
symmetry breaking with no fundamental Higgs boson.  The experimental
signature is the
presence of strongly interacting dynamics at the TeV scale, which
produces new
resonances similar to those found in the hadronic spectrum at the GeV scale.

Although the mechanism for generating electroweak breaking in
technicolour is very
elegant, several difficulties have prevented the
construction of a
fully realistic model.  The first problem is the communication of
electroweak
breaking to the quark and leptonic sectors of the theory.  This can be done
via new
interactions, called extended technicolour (ETC) forces \cite{etc}, which 
couple
quarks to
techniquarks.  If the ETC symmetry is broken (possibly by some dynamical
mechanism)
at a scale $M_{ETC}$
larger than $\Lambda_{TC}$, quarks and leptons receive masses of
the order of
\begin{equation}
m_f \sim \frac{\langle F \bar F \rangle}{M^2_{ETC}} \sim
\frac{\Lambda^3_{TC}}{M^2_{ETC}} ~,
\end{equation}
where $\langle F \bar F \rangle$ is the corresponding technifermionic
condensate.
The trouble is that measurements of FCNC
processes
generally impose stringent lower bounds on $M_{ETC}$, of the order
of 100~TeV.
This means that the ETC mechanism can generate the masses for the first
generation of
fermions, but has difficulties to explain the larger masses of the
second and third
generations.  The task is particularly arduous for the top quark, since
a dynamical
mechanism which explains the large isospin breaking in the difference
between
$m_t$ and $m_b$ generally
leads to large corrections to the $\rho$
parameter, the
ratio between the strengths of the
neutral and charged weak currents.  Finally, the
effect of the
strong technicolour dynamics always gives sizeable corrections to the
electroweak
precision data in LEP1, which have been shown to agree with the Standard
Model with
great accuracy \cite{ew}.

The hope is that these problems can be cured in technicolour theories
with dynamics
substantially different from a scaled-up QCD.  There has been some effort in
this
direction, trying to construct theories in which the ultraviolet
behaviour of the
technifermion self-energy enhances the quark mass contribution, while
the infra-red
behaviour determines the $W$ mass.  This may occur in theories with
slowly running
coupling constants (the so-called walking technicolour \cite{walk}) 
or in fixed-point
gauge
theories \cite{fix}, although the 
non-perturbative nature of the problem prevents
us from making reliable
calculations.

\def\ijmp#1#2#3{{\it Int. Jour. Mod. Phys. }{\bf #1~}(19#2)~#3}
\def\pl#1#2#3{{\it Phys. Lett. }{\bf B#1~}(19#2)~#3}
\def\zp#1#2#3{{\it Z. Phys. }{\bf C#1~}(19#2)~#3}
\def\prl#1#2#3{{\it Phys. Rev. Lett. }{\bf #1~}(19#2)~#3}
\def\rmp#1#2#3{{\it Rev. Mod. Phys. }{\bf #1~}(19#2)~#3}
\def\prep#1#2#3{{\it Phys. Rep. }{\bf #1~}(19#2)~#3}
\def\pr#1#2#3{{\it Phys. Rev. }{\bf D#1~}(19#2)~#3}
\def\np#1#2#3{{\it Nucl. Phys. }{\bf B#1~}(19#2)~#3}
\def\mpl#1#2#3{{\it Mod. Phys. Lett. }{\bf #1~}(19#2)~#3}
\def\arnps#1#2#3{{\it Annu. Rev. Nucl. Part. Sci. }{\bf #1~}(19#2)~#3}
\def\sjnp#1#2#3{{\it Sov. J. Nucl. Phys. }{\bf #1~}(19#2)~#3}
\def\jetp#1#2#3{{\it JETP Lett. }{\bf #1~}(19#2)~#3}
\def\app#1#2#3{{\it Acta Phys. Polon. }{\bf #1~}(19#2)~#3}
\def\rnc#1#2#3{{\it Riv. Nuovo Cim. }{\bf #1~}(19#2)~#3}
\def\ap#1#2#3{{\it Ann. Phys. }{\bf #1~}(19#2)~#3}
\def\ptp#1#2#3{{\it Prog. Theor. Phys. }{\bf #1~}(19#2)~#3}

\end{document}